# Uniaxial Strain-Induced Anisotropic Charge Transfer in Contact Electrification

Yuliang Chen[1,2+]*, Ying Zhang[1,3+], Liang He[1], Steven L. Zhang[1]

[1]School of Material Science and Engineering, Georgia Institute of Technology, Atlanta 30332, United States

[2]National Synchrotron Radiation Laboratory, University of Science and Technology of China, Hefei 230029, China

[3]Key Laboratory of Thermo-Fluid Science and Engineering, Ministry of Education, Xi'an Jiaotong University, Xi'an 710049, China

[+]These two authors contributed equally to this work

Corresponding Authors: ychen3027@gatech.edu

**Abstract**: Even though contact electrification has been studied for a long time, the mechanism of charge transfer still remains elusive. Most of previous reports only focus on the driving force of charge transfer. However, to better understand the mechanism of charge transfer, we believe contact itself for supplying transferring path that charges taking also need to be understood. Here, we focus on the in-plane symmetry of contact in contact electrification by utilizing a uniaxial strain to change the material's isotropic nature to anisotropic. A clear anisotropic charge transfer is observed by contacting axially stretched rubber films at different rotational angles, which could arise from fluctuation of contacting area in microscale. A universal ellipse model is also proposed for qualitatively describing the anisotropy of contact regardless of the specific driving force of contact electrification.

# Introduction

Triboelectrification or contact electrification (CE), as a result of charge transfer by contacting, is ubiquitous from as innocuous as a shock on touching a door-knob to as dramatic as a desert sandstorm.[1-2] Despite even the most basic questions are still being debated for fundamental understanding of CE, such as whether the transferred charges species are electrons,[3-4] ions,[5] or bits of material[6] and why charge transfer occurs between surfaces of identical material,[7-8] it still plays a central role in many useful technologies, such as powder coating, xerography, and electrostatic separation.[9-11] Remarkably, recent research shows that CE based triboelectric nanogenerators (TENGs), which could efficiently harvest low-frequency mechanical energy in ambient environment, have potential for being a new portable power source.[12-15]

Considering previous researches about mechanism of charge transfer in CE, most of them focus on driving force of charge transfer, such as work function difference, micro strain, and statistical variations in materials properties.[4, 7, 16-17] However, the complicacy of CE may originate from a number of mechanisms that become relevant under different conditions rather than following a single universal mechanism. After thinking of the processes of CE carefully, we argue the issue of charge transfer essentially involves two aspects: driving force and contact, Figure 1a. The different properties of two different materials could be considered as the driving force (even if do not know what the exact factors are), which supplies intrinsic power for charge transfer. But, charge transfer could not happen unless the two materials contact each other, which means contact is necessary for suppling the path of charge motion. From this standpoint, contact is also very important. However, the effect of contact was often neglected or had driving force and contact as similar in previous researches for understanding mechanism of charge transfer. For example, in the research about CE between two identical materials at different curvatures (Figure 1b),[18-19] the curvature simultaneously changes the properties of materials (driving force) and the contact's shape, from contact between two flat surfaces to curved surfaces. However, the influence of curved contact is omitted in this example.[20] In contrast, increasing the surface roughness concerns about contact more (Figure 1c).[21-22] The contact's type has changed from two smooth surfaces to two rough ones which increases the contacting area. But the truth is underestimated that increasing roughness, e.g., by chemical corrosion, could have altered surface properties a lot. Therefore, the research about contact is one of the most intriguing questions for fundamental understanding of CE.

In this report, we show the importance of in-plane symmetry of contact. Because strain has been proved as a vital factor for charge transfer between in same material, even reversing the direction of charge transfer,[17-18] two axially stretched strains are applied to two rubber films, which break the in-plane symmetry of the original film without strain.[23] The effect of in-plane symmetry of contact can be explored solely by rotating one of the films at different angles since the driving force predetermined by strains is constant at any angle. Then the uniaxial strain-induced anisotropy of charge transfer was observed. Two models, contacting area in microscale and ellipse mode, were proposed to explain the experimental phenomena.

# Results and Discussion

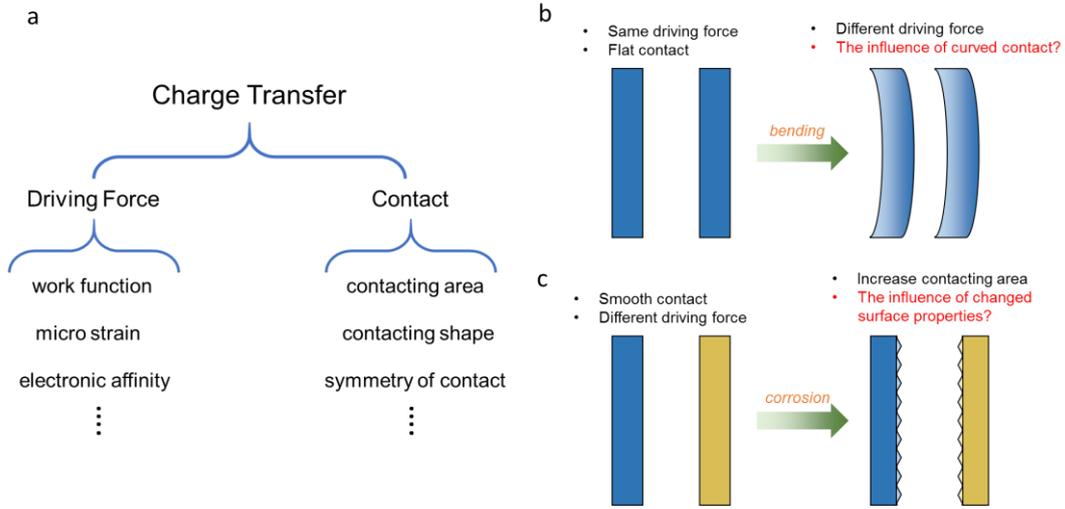

Figure 1. (a) Charge transfer involves two aspects: driving force and contact. (b-c) Two examples of understanding driving force and contact. (b) Focusing on work function (driving force) while neglecting contacting shape in the research of bending identical materials.[18] (c) Focusing on contacting area rather than changed surface properties (driving force) in the research of roughening surfaces.[22]

Figure 2a shows a brief schematic of experimental setup. Two same natural rubber films with different sizes are axially stretched at a nominally same rate. The strain is defined as

$$\varepsilon_- = \frac{d'-d}{d} \times 100\% \quad (1)$$

where $d$ ($d'$) is the length of mark on the rubber films without (with) strain (details in Experimental Section).[17] The subscript "-" means the strain is uniaxial. Here after, all strain is uniaxial except special indication. In Figure 2b, one of the films can be rotated to a targeted angle, and then the amount of charge transfer ($Q$) can be measured by contacting and separating[24] the two films at every angle ($\theta$), for example, in Figure 2c, $\theta = 0°$ (90°) when the strain directions are parallel (orthogonal). The films are in different sizes, so while the film is rotating, the contacting area between the films remains same in macroscale. Usually, the more distinct strain difference between identical materials could improve the amount of charge transfer;[18] however, the giant amount may mask the subtle anisotropy of charge transfer depending on angles. This is the reason the nominally same stretched strains were used. More discussion is in Supplementary Note 1 and Experimental Section.

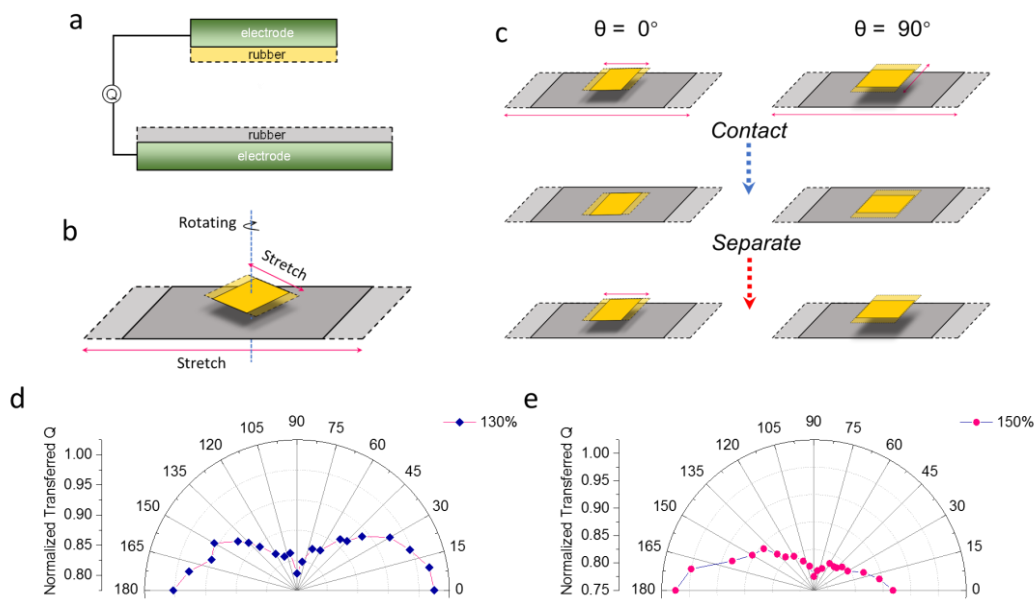

Figure 2. (a) Brief setup to measure the charge transfer by contacting two axially stretched natural rubber films at different rotational angles, as shown in (b). The shape surrounded by solid line is the original rubber without strain, and the shape surrounded by dashed is the rubber under strain. (c) Contact-separation when the stretched directions are parallel ($\theta = 0°$) and orthogonal ($\theta = 90°$). In (b) and (c), the electrodes are not drawn. (d-e) The measured results of charge transfer ($Q$) vary with angle at different stretched strains, $\varepsilon_- = 130\%$ (d), $150\%$ (e).

The results in Figures 2d, e, and S1a, b show a radial plot of the measured transferred charge at different angles, respectively. The anisotropy of charge transfer appearing in films with $\varepsilon_- = 130\%$ and $150\%$ is shown clearly, i.e., $Q$ gradually reduces when $\theta$ changes from 0° to 90°, and then increase from 90° to 180°. Even though the results in these figures are not perfectly symmetric with respect to $\theta = 90°$, they still indicate the charge transfer will be suppressed when the strain directions of two films are orthogonal. In other films with $\varepsilon_- = 175\%$, similar results were observed as shown in Figures S1c and 1d. In contrast, the results of original films without strain ($\varepsilon_- = 0$) are random, which do not show any trend (Figure S2a). Moreover, two isotropic strains ($\varepsilon_o = 70\%$, the superscript "o" means the strain is uniform along all directions is applied rather than a uniaxial strain $\varepsilon_-$) were respectively applied to two round rubber films in different sizes, respectively, whose results indicate that the strain changes the behavior of charge transfer; nevertheless, no clear anisotropy can be observed (Figure S2b) as well. These different behaviors compared with the results in Figures 2d, e and S1a, b indicate that uniaxial strain results in anisotropy of charge transfer. It should be indicated that such anisotropy showing clear trend like Figure 2 is not general, because the different anisotropy in some films with $\varepsilon_- = 100\%$ and $200\%$ were also observed (Figure S3). Although the clear trends cannot be extracted from these anisotropic results, they are also definitely different from the random results of films $\varepsilon_- = 0$ and $\varepsilon_0 = 70\%$.

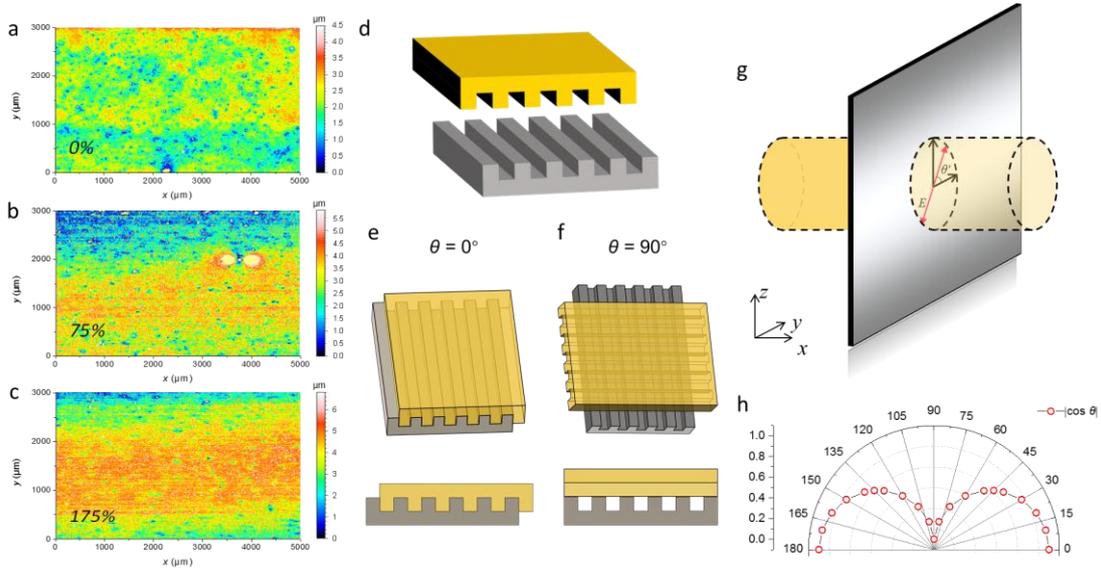

Figure 3. (a-c). The surface morphology of film at different stretched strains, $\varepsilon_- = 0$ (a), 75% (b) and 175% (c). (d) The charge transfer of contact electrification (CE) between two stretched films could be described as two groups of stripes' contact. (e-f) The parallel arrangement (e) and vertical arrangement (f). In these two figures, the top panel is 3-dimensional view and the bottom panel is side view. (g)The schematic of linearly polarized light transmitting a polarizer, which is used to compare the contacting area in microscale between two stretched films. (h) The curve of $|\cos\theta|$.

In order to reveal the secret of the anisotropic charge transfer, we are attempt to explain it from contact rather than driving force, because the strains are constant at every $\theta$. The contacting area ($S$) in microscale between two stretched films may change by rotation although the macro-contacting area is constant at all angels, which is reasonable since many reports have proved that increasing surface roughness promotes charge transfer,[20, 22, 25] namely

$$Q \propto S \quad (2)$$

The measured surface morphology of film at different $\varepsilon_- = 0$, 75% and 175% are in Figure 3a-c, respectively. In Figure 3a, the peaks and valleys in micrometer scale are randomly distributed on the surface of film $\varepsilon_- = 0$, which matches the random results in Figure S2a. In contrast, the interesting thing is that lots of stripes appear in stretched films along the direction of uniaxial strain (x-direction) in Figure 3b ($\varepsilon_- = 75\%$) and c ($\varepsilon_- = 175\%$), which shows two-fold rotational symmetry as the results in Figures 2d and e. These correspondences in symmetry imply the stripes could be responsible for the anisotropic charge transfer of CE between two stretched films.

A phenomenological mode is established for the micro-contacting area with rotation based on two groups of stripes in Figure 3d. Intuitively, like playing lego, the parallel arrangement is more favorable for making two groups of stripes insert each other resulting in maximum contacting area (Figure 3e); the vertical arrangement has the worst effect (Figure 3f); other arrangements are in between. A quantitative description of the contacting area at different angles is difficult since many factors influence the real contact of the stripes in Figures 3b and c, such as shape, Young Modulus. However, the qualitative results can be obtained by comparing with a linearly polarized light transmitting a polarizer in Figure 3g. The polarized direction of polarizer is along y-direction and the light transmits the polarizer along x-direction. $\theta'$ is the angle between electric filed vector ($E$) and y-direction. The intensity ($I$) of a linearly polarized light after passing a polarizer meets formula

$$I = I_0 \cos^2 \theta' \quad (3)$$

$I_0$ is the pristine intensity of the linearly polarized light before passing the polarizer. Because $I_0$ is constant,

$$I \propto |\cos \theta'| \quad (4)$$

In this example, the linearly polarized light and polarizer show the same two-fold rotational symmetry as the stripes. Moreover, the Equation (3) means that it is easiest (most difficult) for light transmitting polarizer when their directions are parallel

(vertical), which is consistent with anisotropic charge transfer. So the qualitative results could be

$$S \propto |\cos\theta| \quad (5)$$

According to Equations (5) and (2),

$$Q \propto |\cos\theta| \quad (6)$$

The curve of $|\cos\theta|$ is plotted in Figure 3h. Indeed, the $Q$ shows the same trend as our experimental data indicated in Figure 2d, e.

Some abnormal area is also measured in Figure 3a-c, which could be defects, impurities, even the exfoliated rubber fragments. We argue that the abnormal area may cause the deviation away from the perfect symmetry comparing Figures 2 and S1 with Figure 3h. Furthermore, more abnormal area may ultimately eliminate the clear trend of anisotropy as the results shown in Figure S3.

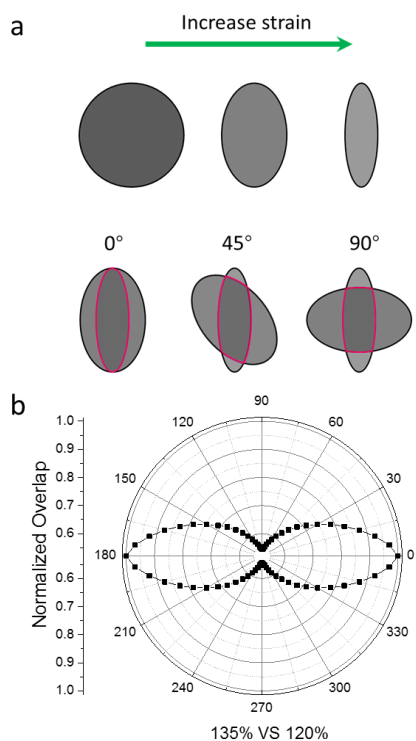

Figure 4. (a) Top panel: An ellipse mode for stretched film. Bottom panel: The overlap of two ellipses surrounded by pink curve is used to qualitatively describe the transferring path of charge transfer between two stretched films. (b) The calculated results of overlap of two ellipses, which are respectively corresponding to $\varepsilon_- = 135\%$ stretched film and $\varepsilon_- = 120\%$ stretched film.

The anisotropic charge transfer substantially roots in the in-plane symmetry of surface morphology. Furthermore, a question arose naturally - if the surface is still smooth after stretching a rubber (or other materials) film whose pristine surface is perfectly smooth, how does the charge transfer change with rotating? This question is pretty interesting, that only thing concerned is that how the symmetry breaking induced by uniaxial strain influences the contact for charge transfers rather than what be exactly transferred in the processes of CE no matter electrons, ions, or bits of material.

Here we come up with an ellipse mode (Figure 4a) based on symmetries. An ellipse is proposed rather than a sphere, because the film's thickness is much smaller than length as well as width. Usually, the film without strain, especially polymer materials, shows isotropy properties, which can be represented by a circle, whose radius ($r$) is

$$r \propto \frac{1}{d} \quad (7)$$

The circle is symmetric at any rotational operation. Then a uniaxial strain is applied for the film, which results in new two-fold rotational symmetry. Thus, an ellipse could be used to describe the changed properties, whose long axis ($a$) and short axis ($b$) are respectively

$$a \propto \frac{1}{d} \quad (8)$$

$$b \propto \frac{1}{d'} \quad (9)$$

From Equations (7) and (8), the length of $a$ is equal to $r$. Based on Equation (9), a stronger strain results in shorter $b$ as shown in the top panel of Figure 4a. It should be emphasized the circle and ellipses are nonobjective, which show the symmetry of material's properties; for example, in a specific system, it could be distribution of chemical potential, micro strain, elementary composition, and the roughness of stripes in Figures 3a-c or comprehensive effect from some of them.

For charge transfer between two stretched films while rotating, an intuitionistic and qualitatively correspondence is the overlapped area of two ellipses at different angles $\theta$ (the area surrounded by pink line in the bottom panel of Figure 4a). The evolution of overlapped area of two ellipses is a complicated mathematic question which is beyond the scope of this work,[26] and is actually difficult to get an analytical solution. The numerical solutions in different situations show in Figures 4b and S4. All results suggest that the overlapped area is minimum when two ellipses are vertical ($\theta =$ 90°), and maximum overlapped area appears at $\theta = 0°$ (obviously, the big ellipse surrounds whole of the small ellipse, bottom panel of Figure 4a).

It also gets a glance based on this model that it could be easier to observe the anisotropy of charge transfer for two films subjected to nominally same strain as indicated at beginning and Supplementary Note 1. In Figure 4b, the minimum overlap at $\theta = 90°$ is 54.4% when $\varepsilon_- = 135\%$ versus $\varepsilon_- = 120\%$ stretched films are used, in contrast, 72.5% for $\varepsilon_- = 60\%$ versus $\varepsilon_- = 100\%$ strain, 72.9% for $\varepsilon_- = 60\%$ versus $\varepsilon_- = 167\%$ in Figure S4. This results also hint that the anisotropy of charge transfer could be harder to be observed for two different kinds of material in CE, since giant charge transfer may hide the subtle anisotropy. Also, the results of this model still support the results in Figures 2 and S1.

## Conclusion

In conclusion, the clear anisotropy of charge transfer is observed by contacting axially stretched rubber films at different rotational angles, which suggests that in-plane symmetry of contact is one of the most important factors of CE. Two proposed models, contacting area in microscale and ellipse mode, qualitatively describe and prove anisotropy of charge transfer, which is accordant with the experimental phenomena. Our results make us understand CE better.

## Experimental Section

*Fabrication of Devices*: The natural rubber films purchased from McMaster-Carr, and the detailed product information is indicated in Table S1. The rubber films can be tailored according to the expected sizes. The original widths of the two films for CE are 57 mm and 15 mm, respectively. For an individual film, its two short edges are installed on each holder, respectively. The stretched degree can be adjusted by changing the distance between holders. The copper electrodes pasted on acrylic boards are beneath of the rubber films. The size of electrode is 45 mm × 45 mm for large film and 13 mm × 20 mm for small one. Because of strains, the films contact with electrodes tightly. The schematic is in Figure S5. In charger transfer measurements, the centers of the two electrodes are aligned in height direction. Diagonal line of the small electrode is 23.8 mm far less than edge length 45 mm of the large electrode, which means the contacting area in macroscale for CE is equal to area of the small electrode at any angle (the macro contacting area is constant). The fabrication of round devices for uniform stretching in all directions is similar to the above processes. The diameter of the round electrodes is 32 mm and 54 mm, respectively. In order to parameterize stretched degree, a line mark along stretched direction (a circle mark on round film) is drawn in advance on film, so the parameter $\varepsilon_-$ ($\varepsilon_o$) can be calculated by Equation (1).[17]

*Characterization*: All the experiments were measured in a glove box with an ultra-pure nitrogen environment (Airgas, 99.999%). The environmental condition was fixed at 20 ± 1℃, 1 atm with additional about 1~1.5 inch $H_2O$ and 0.43% RH. After all films were washed for 20 min by ultrasonication with isopropyl alcohol, distilled water, respectively, and dried by nitrogen blowing, the devices were put in glove box immediately. Before starting measure, the devices were kept in the glove box more than 12 hours. The device with small film and electrode was installed on a stage which was connected with a liner mechanical motor. The device with large film and electrode was installed on a rotating platform. The device could be lifted and pushed down automatically with the help of the linear mechanical motor to realize CE with the other device. The angle can be controlled by rotating platform. The surface level of two devices was carefully adjusted by a gradienter before measurements. The charge transfer was measured by a Keithley 6514 system electrometer. A LabVIEW software platform can achieve real-time data acquisition and analysis. The surface morphology was measured by Nanovea chromatic confocal optical profilometer, lateral resolution: 1.7 μm, z resolution: 8 nm. In order to calculate the numerical solutions of two ellipses' overlap, first we used PowerPoint 2016 to draw the overlap of an angel. Then the picture was opened in Photoshop CS6. The normalized area of overlap can be calculated by reading the amount of pixel.

## Acknowledgments

Y. L. C. and Y. Z. contributed equally to this work. The authors acknowledge support from 111 project (B16038). Y.L.C. thanks China Scholarship Council for supplying oversea scholarship (201706340019).

## Conflict of Interest

The authors declare no competing financial interest.

# Supporting Information

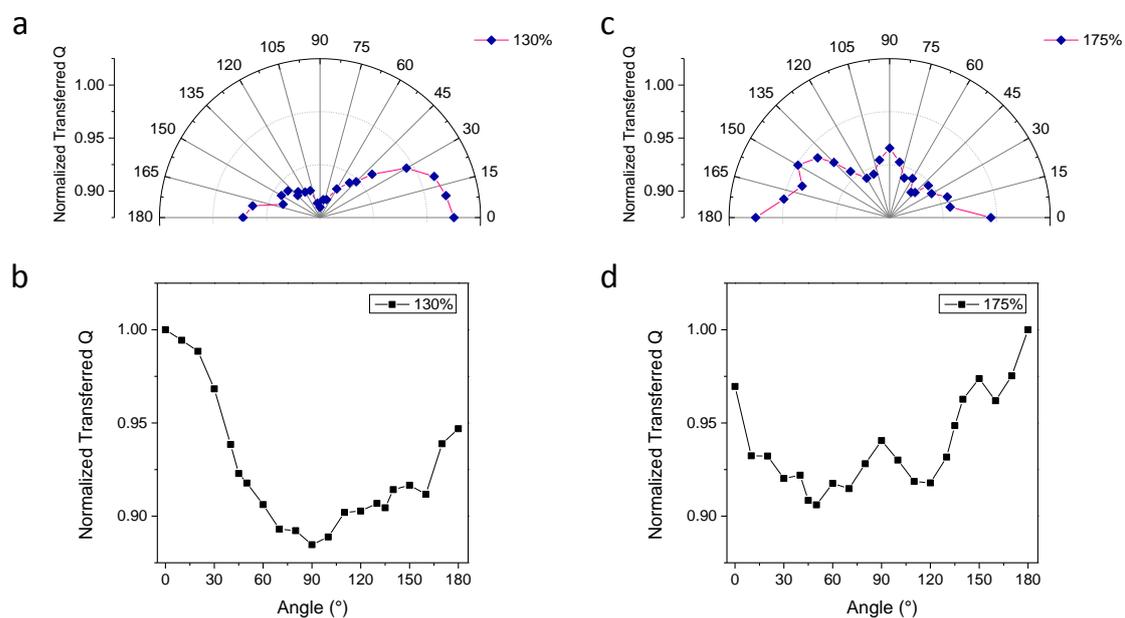

Figure S1. The measured results of charge transfer vary with angle at different stretched strains, $\varepsilon_-$ = 130% (a-b), 175% (c-d). The data is plotted in polar coordinates (top panel) and Cartesian coordinates (bottom panel).

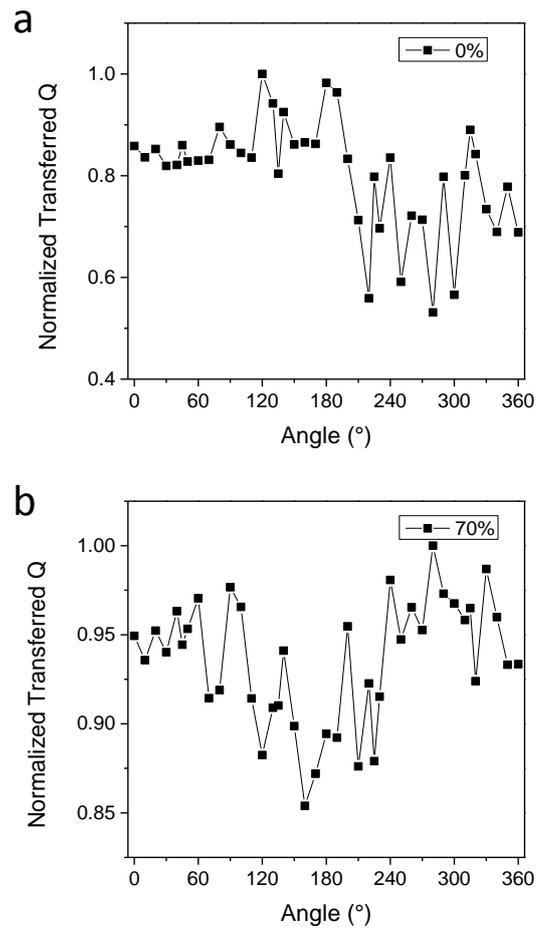

Figure S2. (a) Without strain, the measured results of charge transfer vary with angle. (b) The measured results of charge transfer vary with angle when a $\varepsilon_o$ = 70% uniform streatced strain along all directions is appied to round rubber films.

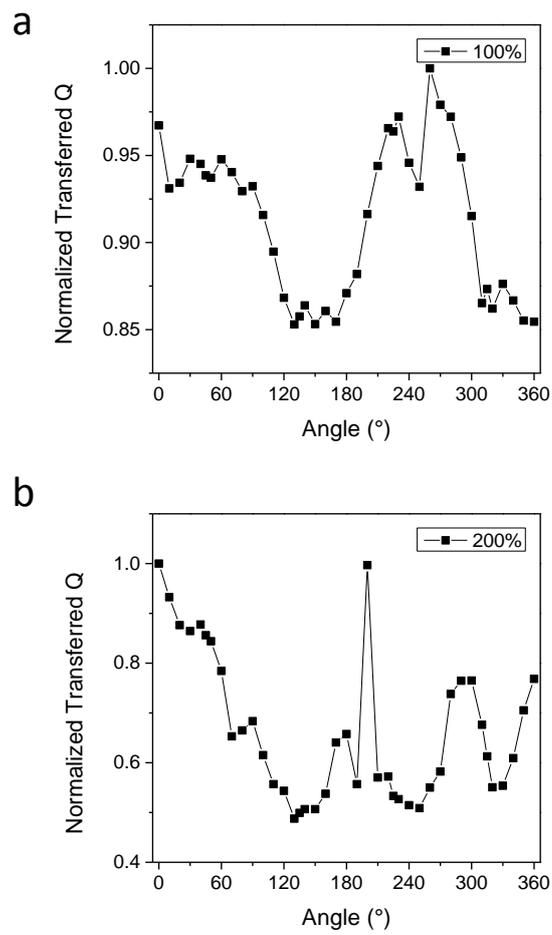

Figure S3. The measured results of charge transfer vary with angle at different stretched strains, $\varepsilon_- $ = 100% (a), 200% (b).

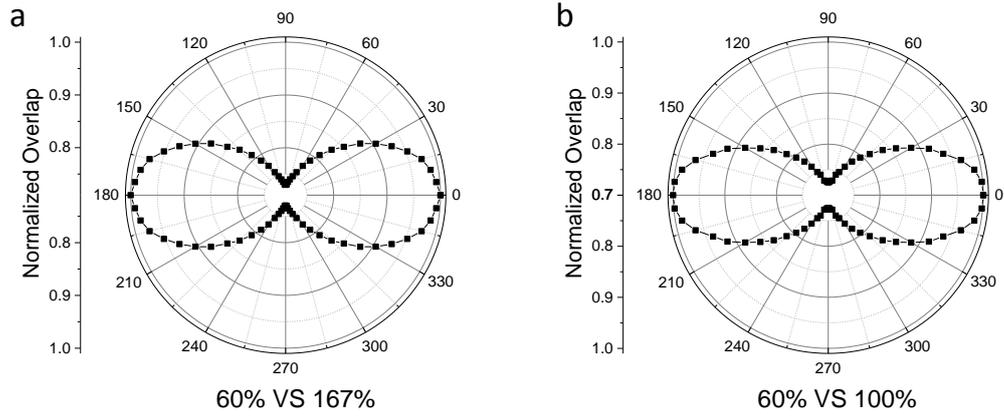

Figure S4. The calculated results of overlap of two ellipses, which are respectively corresponding to $\varepsilon_- = 60\%$ stretched film and $\varepsilon_- = 167\%$ stretched film in (a) and respectively corresponding to $\varepsilon_- = 60\%$ stretched film and $\varepsilon_- = 100\%$ stretched film in (b).

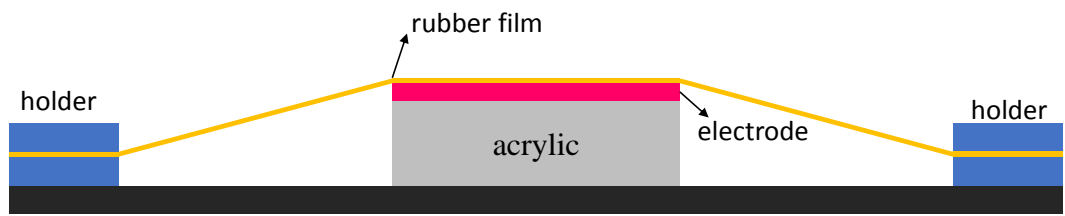

Figure S5. The side view of schematic. Because the electrode is higher than holder, the stretched film contact electrode tightly. Moreover, except the part of film on electrode, the other part cannot take part in CE.

Table S1. The product information of rubber.

| | |
|---|---|
| Construction | Solid |
| Cross Section Shape | Rectangle |
| Material | Natural Rubber |
| Texture | Smooth |
| Thickness | 3/16" |
| Thickness Tolerance | -0.031" to +0.031" |
| Width | 36" |
| Width Tolerance | -1.000" to +1.000" |
| Backing Type | Plain |
| For Use Outdoors | No |
| Temperature Range | -20° to 140° F |
| Tensile Strength | 3,000 psi |
| Color | Tan |
| Specifications Met | ASTM D2000 AA, Made of FDA-Listed Material for Use with Food and Beverage |
| Durometer | 40A (Medium Soft) |
| Durometer Tolerance | -5 to +5 |
| Length Tolerance | -0.5" to +1" |

# Supplementary Note 1

The **nominally** same strains are applied to two rubber films in order to avoid the influence of giant amount of charge transfer when strains with big difference are applied. However, the charge transfer cannot happen between two **absolutely** identical stretched films in principle. In fact, the nominally same strains have subtle difference. When the devices with different sizes films were fabricated, the subtle difference of strains was inevitably introduced into our systems. For example, for nominal $\varepsilon_- = 130\%$, the small size film has $\varepsilon_- = \sim 120\%$ and $\sim 135\%$ for the big one in actual situation. This is also the reason that there are two different ellipses in Figure 4a. But we still use the nominal $\varepsilon_-$ to label each experiment because it will not influence our discussion.